\documentclass[twocolumn]{aastex62}

\usepackage[utf8]{inputenc}
\usepackage{amsmath}

\usepackage{color}

\graphicspath{{./}{figures/}}

\received{}
\revised{}
\accepted{}
\submitjournal{ApJL}

\shorttitle{Charge driven growth}
\shortauthors{Teiser et al.}

\begin{document}

\title{A Smoking Gun for Planetesimal Formation: Charge Driven Growth into a New Size Range}

\correspondingauthor{Jens Teiser}
\email{jens.teiser@uni-due.de}

\author{Jens Teiser}
\affil{University of Duisburg-Essen \\
Faculty of Physics\\
Lotharstr. 1-21 \\
D-47057 Duisburg, Germany}

\author{Maximilian Kruss}
\affiliation{University of Duisburg-Essen \\
Faculty of Physics\\
Lotharstr. 1-21 \\
D-47057 Duisburg, Germany}

\author{Felix Jungmann}
\affiliation{University of Duisburg-Essen \\
Faculty of Physics\\
Lotharstr. 1-21 \\
D-47057 Duisburg, Germany}

\author{Gerhard Wurm}
\affiliation{University of Duisburg-Essen \\
Faculty of Physics\\
Lotharstr. 1-21 \\
D-47057 Duisburg, Germany}

\begin{abstract} %Maximal 250 Wörter

Collisions electrically charge grains which promotes growth by coagulation. We present aggregation experiments with three large ensembles of basalt beads ($150\,\mu\mathrm{m} - 180\,\mu\mathrm{m})$, two of which are charged, while one remains almost neutral as control system. In microgravity experiments, free collisions within these samples are induced with moderate collision velocities ($0 - 0.2 \,\mathrm{m\,s}^{-1}$). In the control system, coagulation stops at (sub-)mm size while the charged grains continue to grow. A maximum agglomerate size of 5\,cm is reached, limited only by bead depletion in the free volume. For the first time, charge-driven growth well into the centimeter range is directly proven by experiments. In protoplanetary disks, this agglomerate size is well beyond the critical size needed for hydrodynamic particle concentration as, e.g., by the streaming instabilities.  

\end{abstract}


\section{Introduction} \label{sec:intro} % Maximal 3500 Wörter im Paper

The first stage of planet formation is dominated by hit-and-stick collisions between small dust and ice grains at small collision velocities \citep{Wurm1998, Blum2008, Johansen2014, Gundlach2015}. Although this first step is fast and efficient, there are several obstacles, which stop this evolution. With increasing agglomerate size, the relative velocities between the colliding aggregates increase \citep{Weidenschilling1993}. This leads to restructuring and compaction \citep{Weidling2009, Meisner2012}. 

If now two of these compact aggregates collide slowly ($< 1\, \mathrm{m\,s}^{-1}$), they rather bounce off than stick to each other. This has been introduced as the bouncing barrier \citep{Guettler2010, Zsom2010}. Several experiments showed that self-consistent growth indeed comes to a halt at a particle size in the millimeter range \citep{Kruss2016, Kruss2017, Demirci2017}. Slight shifts in aggregate size are possible depending on temperatures or magnetic fields \citep{Kruss2018, Kruss2020c, Demirci2019b} but the bouncing barrier is a robust finding.

Collisions and growth in a protoplanetary disk are governed by the interaction between gas and solids. Hydrodynamic processes therefore have a strong effect on particle evolution. Beyond inducing collisions, they can especially change the local particle concentration \citep{Johansen2007, JohansenYoudin2007, Chiang2010, Squire2018}. If a critical solid-to-gas ratio is reached, the mutual gravity between the solids might lead to the direct formation of a planetesimal \citep{Youdin2005, Simon2016,Klahr2020}. This way, barriers in collisional growth could be prevented.

However, while these drag instabilities can be very efficient, they require a minimum size of solids to be present \citep{Drazkowska2014, Carrera2015}. Typically, they work best for so-called pebbles with the Stokes-number (ratio between the orbital period and gas-grain-coupling time) $St \sim 1$. Depending on the disk model and the location in the protoplanetary disk, this Stokes number typically translates into particle sizes of the order of a decimeter though somewhat smaller sizes might still work  \citep{Yang2017} under certain conditions.
Obviously, to explain planet formation, a severe size gap must be bridged between the millimeter size resulting from the bouncing barrier and the decimeter required for the hydrodynamic processes to work. 

This bridge might be a charge dominated growth phase. Collisions and friction between particles lead to charge separation upon contact \citep{Lacks2011}. For a long time, this was attributed either to different materials in contact (different surface energies) or due to different sizes \citep{Lee2015}. Experiments showed that charge separation also occurs for particles of the same size and material \citep{Jungmann2018}.

While the detailed physical processes are poorly understood, the resulting charge distributions in granular samples are well characterized. For a granular sample with particles of the same size and material, a broad charge distribution is the result. By first glimpse, it is similar to a Gaussian distribution, but with heavier tails \citep{Haeberle2018}. The peak position (mean charge) and the full width at half maximum (FWHM) are typically used to characterize the charge distribution of a granular sample. For multiple collisions of particles of the same size and same material the resulting charge distribution peaks at zero charge \citep{Wurm2019}. Within the scope of this paper, the term ''strongly charged'' then refers to the FWHM of a corresponding charge distribution.

Agitating a granular system for a duration of about 10\,min establishes a charge distribution within the system which does not change significantly when the agitation is continued. It only depends on sample and atmospheric parameters as was shown in experiments with monodisperse basalt beads by \citet{Wurm2019}. Larger beads are charged more strongly (larger FWHM) than smaller beads.

Additionally, the charging of a granular sample strongly depends on the surrounding gas pressure. For a constant granular sample the width of the reached charge distribution follows a curve similar to Paschen's law of electrical breakthrough in gases. The width (FWHM) of the charge distributions reaches a minimum at a characteristic pressure ($100\,\mathrm{Pa}\,-\,\mathrm{few}\,100\,\mathrm{Pa}$), depending on the sample. At larger pressures, the reached width increases gently, while it increases sharply for pressures smaller than the characteristic value.

%Recent experiments showed that the growth is strongly influenced by electrical charges \citep{Steinpilz2020a}. Electrical charges affect the collision dynamics directly \citep{Jungmann2018} and promote the growth of aggregates \citep{Lee2015, Steinpilz2020a}.
%Even mutual collisions between particles of ''identical'' grains, i.e. consisting of the same material and having the same size still lead to a broad charge distribution,  also working in low pressure environments below $p < 100\,\mathrm{Pa}$ as given in protoplanetary disks \citep{Wurm2019}. \JT{Description of charge distributions, Peak position, width,...}. 
%In detail, collisional charging of millimeter particles leads to a complex charge distribution on the particle surface and the interaction between the particles is not only governed by the net charge but also by dipole or higher multipole interactions \citep{Jungmann2018, Steinpilz2020b}. \JT{Here: No discharge, when particles get in contact, including references, discharge timescales if possible}. 
%Discharging has to be considered to counterbalance charging, but at least cosmic radiation of protoplanetary disks, comparable to the radiation level of the upper Earth's atmosphere, does not alter the charge state significantly \citep{Steinpilz2020a, Jungmann2021}. 

\section{Charge driven aggregation and stability}

%\textbf{\subsection{How charges change aggregation}}

It is obvious that two grains of opposite charge attract each other as a first step in aggregation. This effectively changes the collisional cross section. It is not obvious \textit{a priori} though why an initial charge on individual grains should be beneficial for the aggregation process later on, once agglomerates have formed. In fact, once two grains of the same absolute charge, but opposite sign, collide and stick to each other this dimer is overall neutral. The long-range Coulomb interaction of net charges is no longer present and the collisional cross section will no longer be strongly enhanced.

However, insulating grains do not discharge upon contact. This even holds for metal spheres, if their surface is not extremely cleaned from any contamination \citep{Genc2019, Kaponig2020}. A dimer or a more complex aggregate still hold charges on their surface. It has to be noted at this point that collisions lead to charge separation (not neutralization) in the first place and grains charged this way have a complex charge pattern with patches of negative and positive charges on their surface \citep{Grosjean2020, Steinpilz2020b}. This is also valid for grains which are net neutral. The typical configuration are therefore multipoles even on a single grain.

Certainly, there are ways to discharge and neutralize grains, which depends on the environment (water content, gas pressure, temperature, radiation, material conductivity). In the experiments here, discharge takes hours, under protoplanetary disk conditions it might be years \citep{Steinpilz2020a, Jungmann2018, Jungmann2021} (and running experiments by Steinpilz et al., personal communication).

So, while these multipole configurations in aggregates might not attract other grains from far away, charges remain highly important during contact. As Coulomb forces decrease with distance of two charges $r_c$ as $1/r_c^2$, two oppositely charged spots on a surface close to the contact point can dominate the sticking force, independent of the net charge budget of the grains. Therefore, collisions lead to sticking at much higher collision velocities for charged grains \citep{Jungmann2018}. It is important to note that in an ensemble of grains net charge is only a proxy that is easily accessible to confirm that grains have a surface charge pattern. Nevertheless, the multipoles determine sticking forces.

Aggregates are also more stable, i.e. higher collision velocities are required to destroy them compared to uncharged aggregates \citep{Steinpilz2020a}. Charge patches glue aggregates together. A simple analog for this situation is a salt crystal, which is overall neutral but the alternating charges still provide strong attraction. This way, we expect collisionally charged grains to grow far beyond the bouncing barrier. Therefore, there is a high potential in charge driven growth.

It is still unclear though, how large agglomerates can grow this way. In \citet{Steinpilz2020a}, cm-size charged agglomerates were observed but their direct formation from individual grains was not traced and took place prior to the free-floating phase in an agitated granular bed. The effects of charging were shown by means of numerical simulations matching the experiments in that case. The key question thus remains: Is charge driven coagulation able to provide the necessary aggregate sizes for drag instabilities to take over?

\section{Experiment} \label{sec:exp}

To investigate how large agglomerates can form by collisions of small charged particles, a microgravity experiment is currently being developed for sub-orbital platforms. 
%The original goal of the experimental setup was to analyse the formation of agglomerates on the timescale of minutes, using a novel flight opportunity for suborbital flights (MIURA 1, PLD Space). The operational constrains of the flight opportunity (shared payload, scheduled for maiden flight) limit the experiment dimensions and the technical complexity of the experiment. 
Here, we report on the first shorter time microgravity experiments during this development, which were conducted at the Bremen Drop Tower (ZARM).
%, primary to qualify the experiment hardware for the later suborbital flight. 
Using the catapult mode, microgravity with residual acceleration of $< 10^{-6}\,\mathrm{g}$ and a duration of $9.2\,\mathrm{s}$ could be used.

\subsection{Experimental setup}

\begin{figure}
    \centering
    \includegraphics[width = \columnwidth]{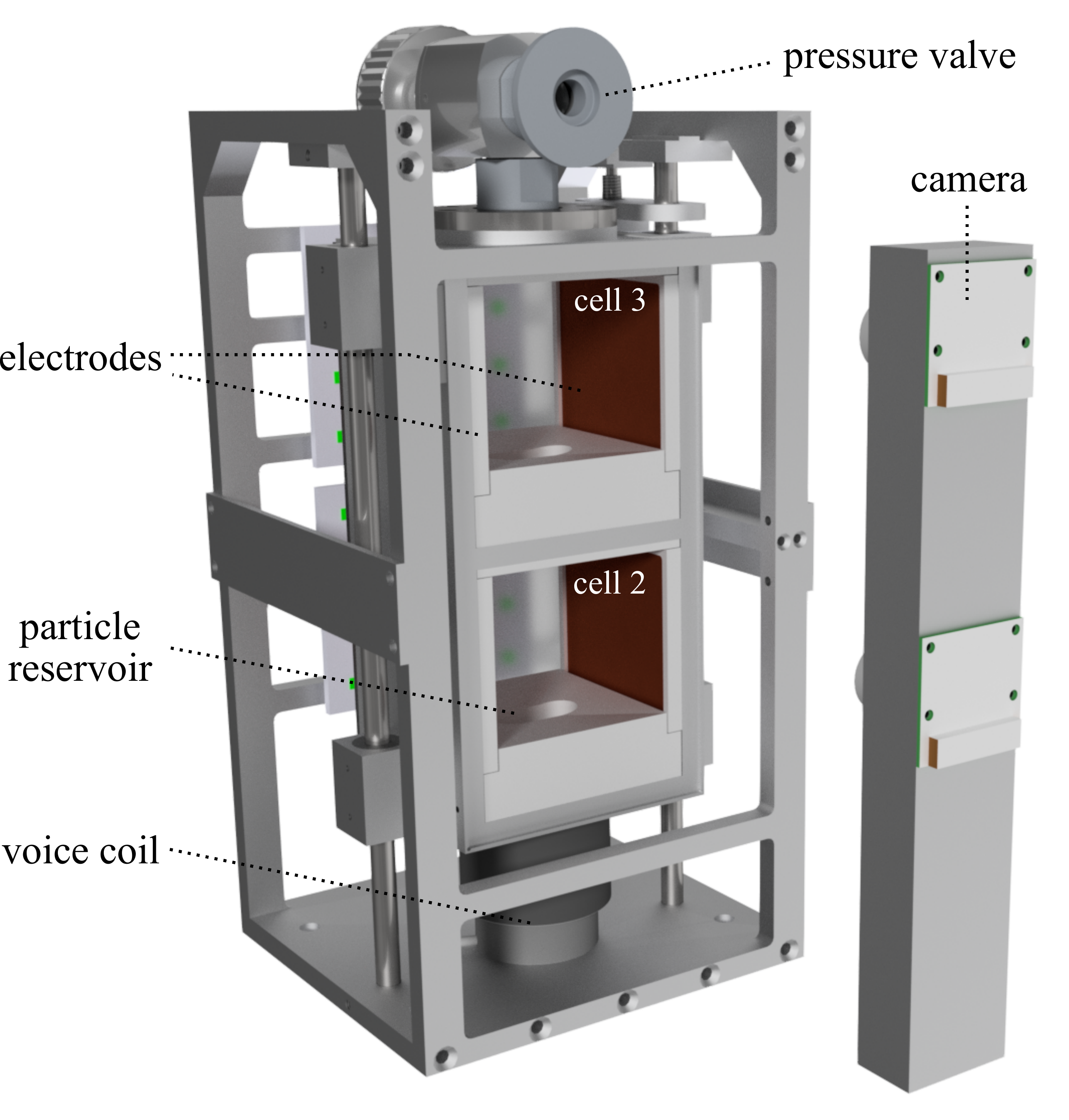}
    \caption{Schematic view of test cells 2 and 3. The upper cell (3) is a vacuum chamber ($p \approx 20\,\mathrm{Pa}$), while cell 2  is at normal pressure.}
    \label{fig:exp}
\end{figure}

The central part of the experimental setup consists of three test cells, two of which are shown in Fig.~\ref{fig:exp}. All cells have the same geometry with a free volume of $50\,\mathrm{mm}\times 50\,\mathrm{mm}\times 42\,\mathrm{mm}$ and an additional particle reservoir of 14\,mm depth and 25\,mm length. While cells 2 and 3 are placed in the same unit, cell 1 is placed in a single unit with half the height. Test cell 3 (the upper one in Fig.~\ref{fig:exp}) is designed as a vacuum chamber with a pressure of $20\,\mathrm{Pa}$, while the other two cells are at normal pressure. 

The side walls of the test cells are copper electrodes which are part of  a capacitor at which a DC-voltage of 4\,kV can be applied.
%$+\,2\,\mathrm{kV}$ on the right hand and $-\,2\,\mathrm{kV}$ on the left hand side. 
The experiments are observed with a Raspberry Pi camera (30\,frames/s, resolution: $1640\,\times 1230\,\mathrm{Px}$) using bright field illumination in a backlight configuration. The optical resolution is of the order of the particle size. 

Each sample consists of basalt beads with a size distribution between $150\,\mathrm{\mu m}$ and $180\,\mathrm{\mu m}$ diameter (Whitehouse Scientific). A total amount of 6\,g is used in each cell. To reduce collisions between basalt beads and different materials, the top and the bottom of each test cell (including the particle reservoir) are coated with the same basalt beads.

\subsection{Experiment protocol}

The test cells can be agitated with a harmonic motion using a voice coil mounted underneath. Prior to the catapult launch this agitation is used to charge the basalt beads for test cells 2 and 3 by collisions and friction due to constant agitation on ground. The duration is 20\,min with a frequency of 14\,Hz and an amplitude of 4.6\,mm (peak to peak), similar to the experiment protocol in \citet{Steinpilz2020a}. During this period the sample mostly remains within the sample reservoir and the beads are exposed to numerous collisions and friction among each other. Even in case the particles leave the reservoir, they are only exposed to particle-particle collisions, as the bottom of the test cells is coated with basalt beads and tilted by an angle of $4^\circ$, so basically no particles hit the side walls. 

The sample was left at rest for about 10\,min between this agitation period and the catapult launch, which does not change the charge distribution significantly. In contrast to cells 2 and 3, test cell 1 was not agitated on ground, so a sample with minimum charge is used as control experiment. In microgravity, the agitation is used to distribute the sample at the beginning and to keep up a sufficient collision rate to see growth on the short timescales available at the drop tower.

The experiment protocol has been changed for the different experiments, so that different aspects of the planned long-duration experiments could be tested on a short timescale. A high voltage at the capacitor plates can be used to estimate the charges in aggregates, but immediately stops further growth processes as the volume is cleared from particles. Agitation can be used to induce a high collision rate, but no charge measurement or particle tracking is possible during this agitation. The different parameters used in the presented experiments are described in section~\ref{sec:results}.

\subsection{Collision velocities}

\begin{figure}[tb]
    \centering
    \includegraphics[width = \columnwidth]{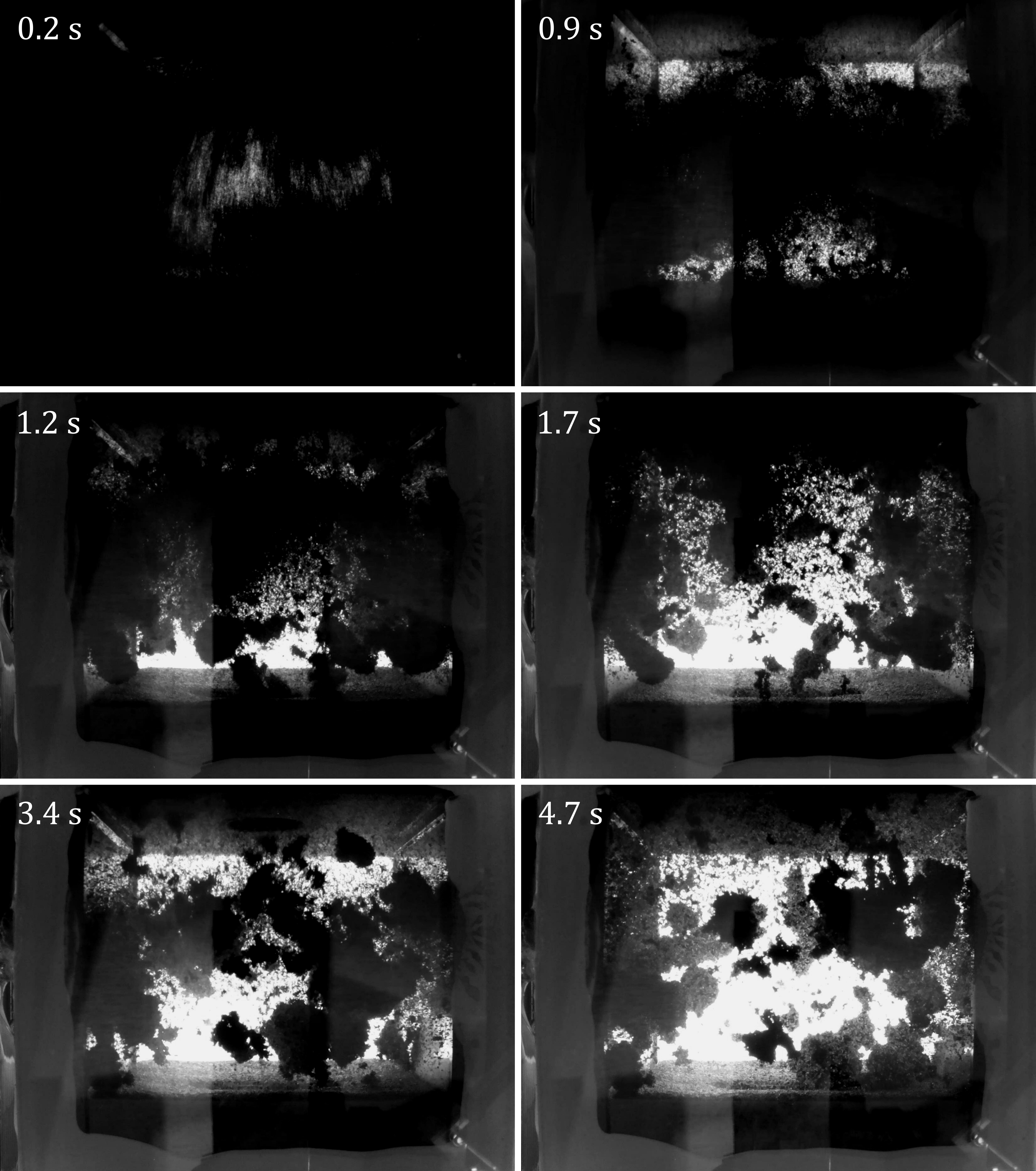}
    \caption{Evolution of the particle ensemble in cell 3 during one experimental run at different times, starting directly after the begin of microgravity and ending when the sample has reached a final state while the test cell is at rest. The width of the chamber of 50\,mm can be used as a scale.}
    \label{fig:flight3}
\end{figure}

Due to the limitations of the optical system and the large particle concentration within the test cells, it is not possible to track single particles. However, the collision velocities can be estimated using the parameters of the agitation cycle. As the agitation follows a harmonic, frequency $f$ and maximum amplitude $A_0$ directly translate in a maximum velocity of the test cells via $v_{\rm{max}} = 2\pi f\cdot A_0$. For the parameters used in Fig.~\ref{fig:flight3} ($f = 14\,\mathrm{Hz}$, $A_0 = 1.2\,\mathrm{mm}$) this results in a maximum velocity of $v_{\rm{max}} = 0.11 \,\mathrm{m\,s}^{-1}$. In case of perfectly elastic collisions between the particles and the experiment walls (top and bottom), a resting particle could get a maximum velocity of $v_{\rm{col}} = 2 v_{\rm{max}}$ or even larger in case of an initial velocity towards the wall. 

Indeed, non-charged basalt beads collide rather elastically with a coefficient of restitution $\epsilon = v_{\rm{after}}/v_{\rm{before}}$ of the order of $\epsilon = 0.9$ \citep{Bogdan2019}. It has to be noted that smaller basalt beads are used and collision velocities are lower in the drop tower experiments compared to the work by \citet{Bogdan2019}. Additionally, this situation is slightly more complicated. As the surfaces of the bottom (including the particle compartment) and the top are coated with the same basalt beads as the sample particles, all collisions are collisions between beads at a random impact parameter. Also the charges of the beads themselves influence the collision behavior \citep{Jungmann2018}, as the coefficient of restitution gets smaller for larger charges.

Altogether we estimate the maximum collision velocities to be roughly the same as the maximum velocities of the test cells. Additionally, collisions in the free volume of the test cell will damp the original velocity distribution.  We therefore assume a broad velocity distribution from almost zero to the maximum velocity of the test cell. 

\section{Results} \label{sec:results}

 %collision velocities

%At 14 Hz: 0.088 m/s max

%At 1 Hz: 0.019 m/s max

%Parameters for exp in fig. \ref{fig:flight3}: 
%ground: 20 min 14 Hz, 10 min rest
%flight: 14 Hz 1 s, HV 1 s, 14 Hz 1 s, afterwards at rest

The experiments presented were planned to qualify an experiment hardware for suborbital flights and to test parts of the experiment protocol. Here, we present three different experiment protocols, which were used to show different aspects of the upcoming long-duration experiments.

\subsection{First growth}

The first experiment protocol was used to check if a rather homogeneous particle distribution can be generated by agitation in microgravity. Here, the sample was agitated with a frequency of 14\,Hz, an amplitude of 1.2\,mm, and a duration of 1\,s, starting at the beginning of the microgravity phase. After a short break of 1 s, the agitation was then repeated for a duration of $1\,\mathrm{s}$. Afterwards, the test cells were not moved until the end of the microgravity.

%Figure \ref{fig:flight3} shows two states of the sample distribution for this experiment, the initial distribution of the basalt beads (after the last agitation run) and the final distribution in all three test cells. Test cell 1 with the minimally charged sample (Fig.~\ref{fig:flight3}, top panel, no agitation on ground) can be clearly distinguished from the other two samples. Only very few agglomerates exceed the size of 1\,mm, while most of the particles build very small entities or do not even agglomerate.

Fig.~\ref{fig:flight3} shows the temporal evolution of the particles in the vacuum chamber (cell 3) for this experiment. It starts directly after the last agitation cycle from a well distributed sample, which blocks the illumination almost completely. While the test cell is kept at rest a clustering process can be observed. After around 5\,s, the final state is reached as the grown clusters do not collide anymore.

Of the three cells, the vacuum chamber shows the most homogeneous sample distribution in the initial state and the largest agglomerate sizes in the final state (see also Fig.~\ref{fig:result}). According to \citet{Wurm2019} basalt beads charge differently depending on the surrounding pressure. At pressures of about $100\,\mathrm{Pa\,-\,\mathrm{few}\, 100\,\mathrm{Pa}}$ the charge distribution is narrowest. For lower pressure the width of the charge distribution rises steeply \citep{Wurm2019}. Additionally, particle motion is not damped significantly by gas drag, as the gas-grain coupling time exceeds the experiment duration. Therefore, this directly leads to larger particle velocities and a higher collision rate. However, the maximum size reached in this experiment run is still restricted to a few millimeters. This can be attributed to the declining collision rate, as the particle velocities are damped by collisions and/or gas drag (depending on the test cell) and therefore also the collision probability goes down.

\subsection{Charges}
\label{sec:charges}

The charge distribution of single basalt beads cannot be obtained from the data available, as the spatial and temporal resolution of the camera system are not suitable for this. However, the agitation method and therefore the process of particle charging is almost identical to previous studies, either with glass beads \citep{Jungmann2018, Jungmann2021, Steinpilz2020a} or with basalt beads \citep{Wurm2019}. The width (FWHM) of the corresponding charge distribution scales with the particle size \citep{Wurm2019}, with many studies treating charges on insulators as surface charges only \citep{Lee2018, Grosjean2020, Steinpilz2020b}. With a similar charge density on the surface as in \citet{Wurm2019}, a charge distribution centered at zero charge with a FWHM of about $3\cdot10^{-13}\,\mathrm{C}$ can be expected.

\begin{figure}[tb]
    \centering
    \includegraphics[width = \columnwidth]{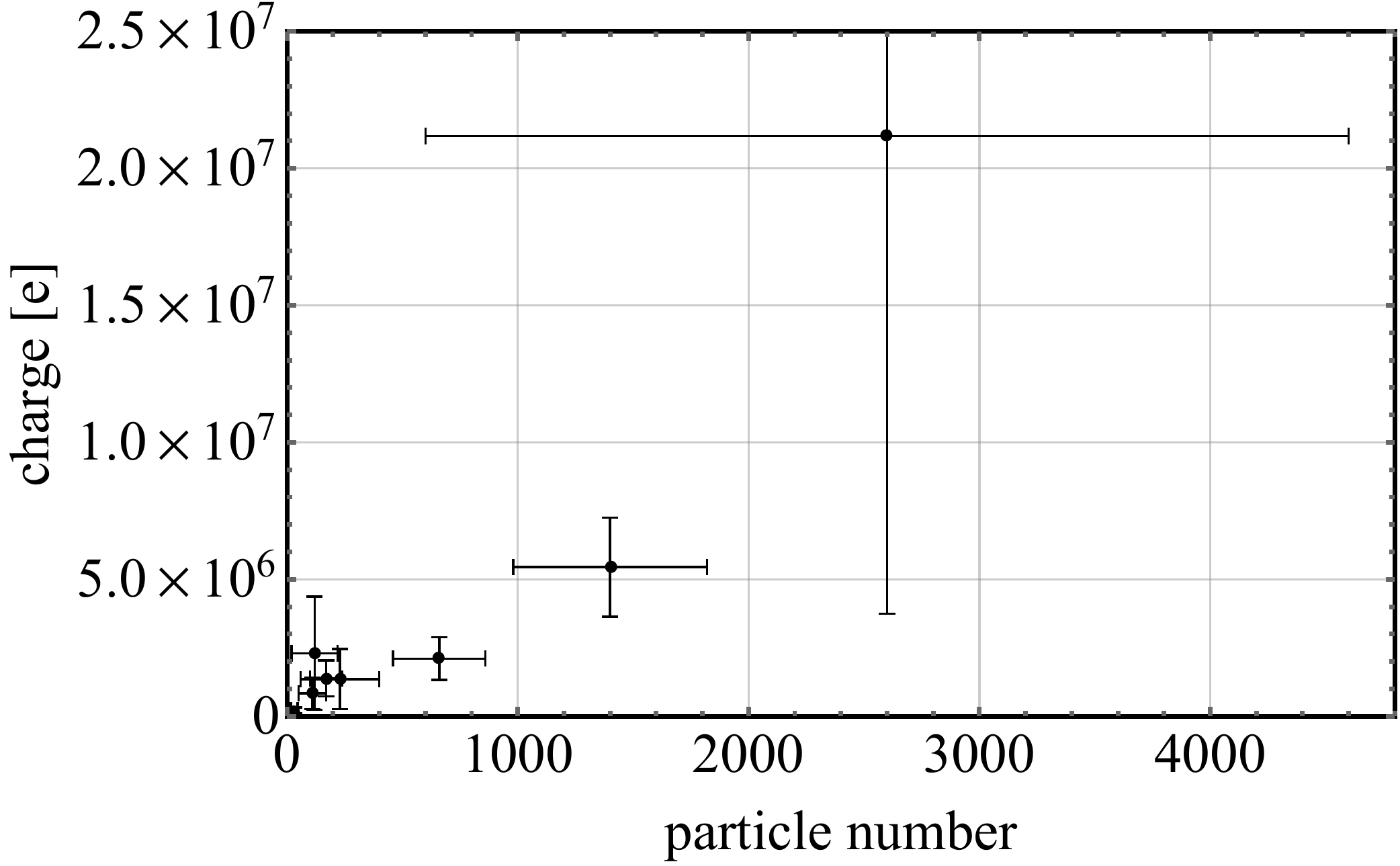}
    \caption{Absolute estimated charges of resulting clusters formed during microgravity. The large error bars result from the high uncertainty of the masses of the agglomerates.}
    \label{fig:charges}
\end{figure}

To roughly estimate the charges on the agglomerates we performed an experiment in which the beads were shaken for 5\,s during microgravity (f = 14\,Hz, A = 1.2\,mm). As shown in Fig.~\ref{fig:result} (middle) cm-sized agglomerates form in cell 2. After agitation, a voltage of $\pm 2\, $kV is applied in that cell which accelerates all charged particles towards the electrodes. These single agglomerates are tracked manually and their acceleration is translated to the amount of charge they carry. For this the mass of the agglomerates and its error is estimated via their cross-sections in the images. Fig.~\ref{fig:charges} shows that the agglomerates are formed from several hundred up to thousands of single particles. Their typical charge is up to $10^7$ electron charges. We note that this is only an estimate of the order of magnitude as a detailed analysis is beyond the scope of this paper and not possible with the available data from the short-time experiments. However, this indicates that there are abundant charges on the clusters which might play a role in the agglomeration process.

\subsection{Growing tall}

%\begin{figure}
%    \centering
%    \includegraphics[width = 8.cm]{/figures/exp10pi3.png}\\
%    \includegraphics[width = 8.cm]{/figures/exp10pi1.png}\\
%    \includegraphics[width = 8.cm]{/figures/exp10pi2.png}
%    \caption{Final particle distributions. Top: sample with minimum charge (no shaking in advance). Middle: 20 minutes shaking in advance, normal pressure. Bottom: 20 min shaking in advance, vacuum (20 Pa).}
%    \label{fig:result}
%\end{figure}

\begin{figure}[tb]
    \centering
    \includegraphics[width = \columnwidth]{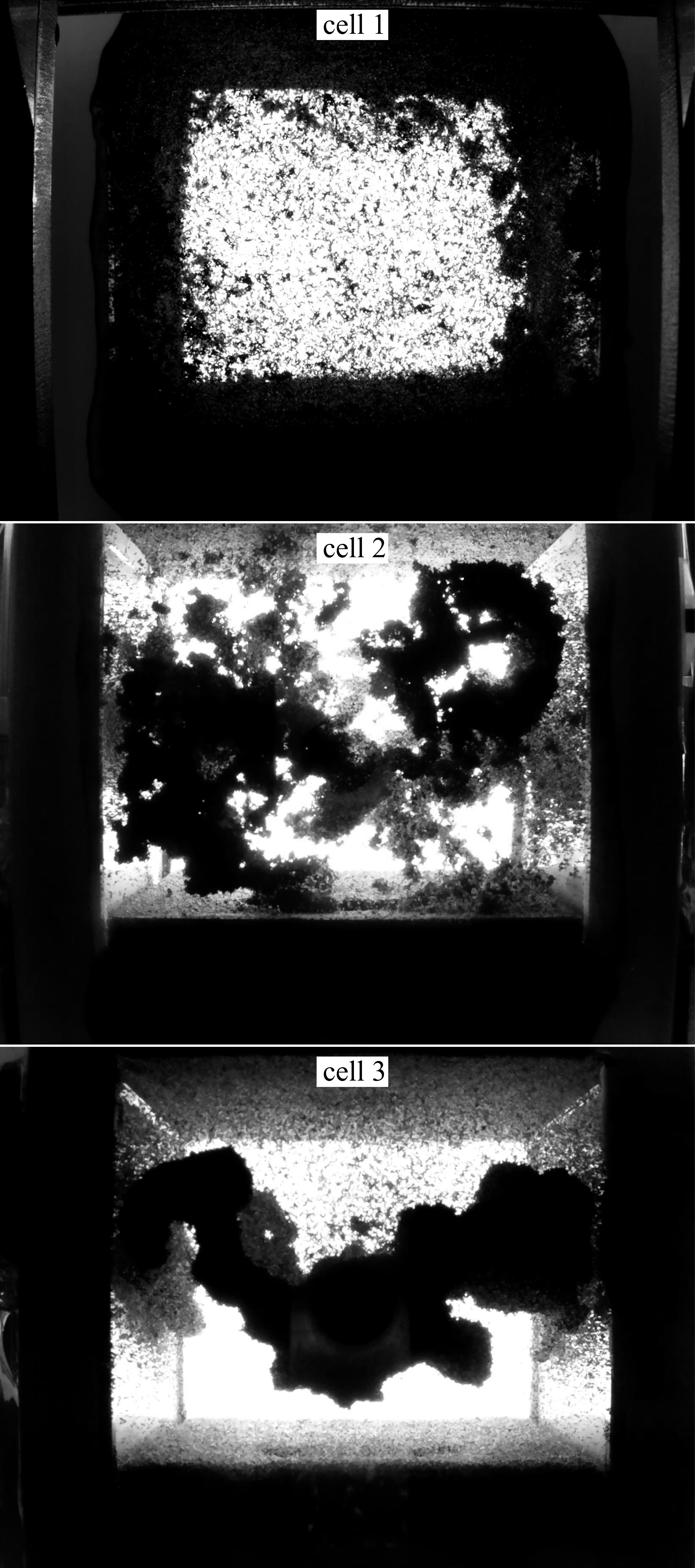}
    \caption{Final particle distributions with continuous agitation of 1\,Hz under microgravity. Top: sample with minimum charge (no shaking in advance). Middle: 20\,min shaking in advance, normal pressure. Bottom: 20\,min shaking in advance, vacuum (20\,Pa).}
    \label{fig:result}
\end{figure}

The evolution presented in Fig.~\ref{fig:flight3} shows that growth by collisions of charged basalt beads is possible in principle. On the other hand it becomes clear that the collision rate is crucial for the outcome and has to be kept on a high level. This was considered in the experiment shown in Fig.~\ref{fig:result}. Here, the experiment protocol was changed to maintain a certain collision rate, while observation of the grains is still possible. With the onset of microgravity, the test cells were agitated with $f = 14\,\mathrm{Hz}$ and $A_0 = 1.2\,\mathrm{mm}$ amplitude for a duration of 3\,s. Afterwards, the agitation frequency was reduced to $f = 1\,\mathrm{Hz}$ for a duration of 4\,s, resulting in an amplitude of 3\,mm and a maximum velocity of $v_{\rm{max}} = 0.02 \,\mathrm{m\,s}^{-1}$. Afterwards, the test cells were at rest for the last 2\,s of microgravity.

Fig.~\ref{fig:result} shows the final particle distributions during this experiment run. It also reveals the systematic differences between the three test cells. The least charged sample (top) only shows minor growth in comparison to the charged sample in the test cell with atmospheric pressure (middle) where larger entities evolve. The vacuum cell (bottom) shows a striking result. Almost all particles are incorporated into one large agglomerate, which therefore has a width of 5\,cm (from wall to wall), a thickness of about 2\,cm and a total mass of about $6\,\mathrm{g}$.\\

The free volume between the larger agglomerates is also of great interest to interpret the result. In the least charged sample, the particles remain widely distributed in the entire volume. Also in the test cell with atmospheric pressure there is still a significant amount of single beads (or only very small agglomerates) in the free volume between the larger aggregates. This is totally different in cell 3, where the maximum charges can be expected. Single particles are either incorporated into the major agglomerate or stick to the walls of the cell. The free volume is almost completely cleared from small particles. Due to this particle depletion the growth process comes to a halt. Therefore, it can be assumed that the final distribution does not show the maximum sizes achievable by such collisions.

\subsection{Stability}

\begin{figure}[tb]
    \centering
    \includegraphics[width = \columnwidth]{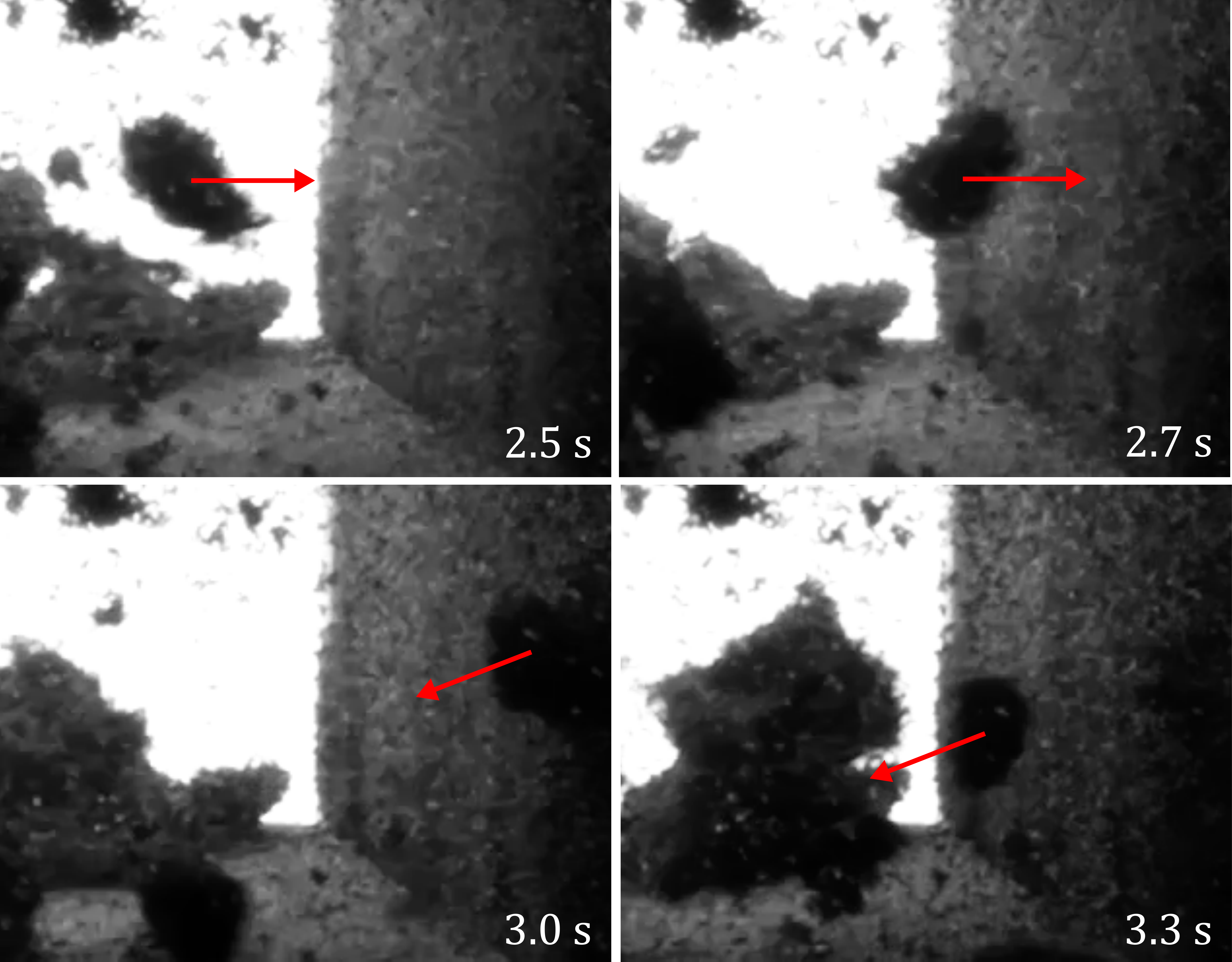}
    \caption{Sequence of a 2.5 mm cluster (average diameter) colliding with the electrode. Its impact velocity is about 2 cm/s. The red arrow shows the direction of motion.}
    \label{fig:collision}
\end{figure}

When applying an electric field some large clusters are accelerated towards the electrodes and therefore reach high impact velocities. This can be used to estimate the stability of these clusters. Similar to chapter \ref{sec:charges} these clusters were tracked manually and their impact velocities determined. An example of a cluster colliding and bouncing off the wall is shown in Fig. \ref{fig:collision}.

Collisions between clusters and the (side) walls occur at impact velocities from $10^{-2}\,\mathrm{m\,s}^{-1}$ to $2.1\cdot 10^{-2}\,\mathrm{m\,s}^{-1}$, while the typical cluster sizes (average diameters) range from $2.5\,\mathrm{mm}$ to $5.3\,\mathrm{mm}$. No fragmentation was observed, only sticking or bouncing were observed. Relative velocities in protoplanetary disks depend on the sizes of the collision partners and are lowest for particles of similar size \citet{Weidenschilling1993}. For equally sized particles, the collision velocities are $\leq 10^{-2}\,\mathrm{m\, s}^{-1}$, even for particles of a few cm in size \citep{Weidenschilling1993}. As a collision with a solid wall is much more severe than mutual collisions between equally sized clusters, mutual collisions in protoplanetary disks will not destroy the grown agglomerates.

The relative velocities are larger for particles of different size, so single particles hitting an agglomerate will be faster. However, the velocities are still in the range of $10^{-1}\,\mathrm{m\, s}^{-1}$, which is exactly in the velocity range of the single basalt beads during the agitation cycles, resulting in growth.

\section{Conclusion and outlook} \label{sec:conclusion}

Charge driven coagulation has already been presented by \citet{Steinpilz2020a} as a possible mechanism to overcome the bouncing barrier in planet formation. Although they could show that the size distribution of agglomerates in an ensemble changes when electrical charges are present, the maximum sizes found were still restricted. Of course, this first step might already be sufficient to help hydrodynamic processes to facilitate further growth \citep{Yang2017, Schaffer2018}. However, this process is still fragile with respect to the corresponding hydrodynamic models.

Here, we present a smoking gun for coagulation to large agglomerate sizes. Although the detailed growth processes are still hidden due to the experimental limitations, there is the proof of concept that small charged grains indeed grow well into the centimeter range and possibly further starting from scratch, i.e. starting with a cloud of individual grains. We observed that growth is only possible if the grains are charged. The size of the largest agglomerates in the experiments presented here is limited by the particle supply (no further collisions) and a non-sufficient experiment duration. Charge driven growth might even continue into the decimeter range. It has to be noted though, that agglomerates in this size range are in danger of destruction by wind erosion, as they already move fast with respect to the surrounding gas. 

Future experiments will enable us to trace the charge driven coagulation in more detail. As the experimental setup is already dedicated for long duration experiments on suborbital flights, a much better understanding of the detailed growth process can be expected.

\acknowledgments

We thank the anonymous reviewer for the fruitful comments, which helped to improve the manuscript. This work is supported by the German Space Administration (DLR) with funds provided by the Federal Ministry for Economic Affairs and Energy (BMWi) under grant 50WM1762.

\vspace{5mm}

\bibliography{references}

\begin{thebibliography}{}
\expandafter\ifx\csname natexlab\endcsname\relax\def\natexlab#1{#1}\fi
\providecommand{\url}[1]{\href{#1}{#1}}

\bibitem[{{Blum} \& {Wurm}(2008)}]{Blum2008}
{Blum}, J., \& {Wurm}, G. 2008, \araa, 46, 21

\bibitem[{{Bogdan} {et~al.}(2019){Bogdan}, {Teiser}, {Fischer}, {Kruss}, \&
  {Wurm}}]{Bogdan2019}
{Bogdan}, T., {Teiser}, J., {Fischer}, N., {Kruss}, M., \& {Wurm}, G. 2019,
  \icarus, 319, 133

\bibitem[{{Carrera} {et~al.}(2015){Carrera}, {Johansen}, \&
  {Davies}}]{Carrera2015}
{Carrera}, D., {Johansen}, A., \& {Davies}, M.~B. 2015, A\&A, 579, A43.
\newblock \url{https://doi.org/10.1051/0004-6361/201425120}

\bibitem[{{Chiang} \& {Youdin}(2010)}]{Chiang2010}
{Chiang}, E., \& {Youdin}, A.~N. 2010, Annual Review of Earth and Planetary
  Sciences, 38, 493

\bibitem[{{Demirci} {et~al.}(2019){Demirci}, {Krause}, {Teiser}, \&
  {Wurm}}]{Demirci2019b}
{Demirci}, T., {Krause}, C., {Teiser}, J., \& {Wurm}, G. 2019, \aap, 629, A66

\bibitem[{{Demirci} {et~al.}(2017){Demirci}, {Teiser}, {Steinpilz}, {Landers},
  {Salamon}, {Wende}, \& {Wurm}}]{Demirci2017}
{Demirci}, T., {Teiser}, J., {Steinpilz}, T., {et~al.} 2017, \apj, 846, 48

\bibitem[{{Dr{\c{a}}{\.z}kowska} \& {Dullemond}(2014)}]{Drazkowska2014}
{Dr{\c{a}}{\.z}kowska}, J., \& {Dullemond}, C.~P. 2014, \aap, 572, A78

\bibitem[{Genc {et~al.}(2019)Genc, Mölleken, Tarasevitch, Utzat, Nienhaus, \&
  Möller}]{Genc2019}
Genc, E., Mölleken, A., Tarasevitch, D., {et~al.} 2019, Review of Scientific
  Instruments, 90, 075115.
\newblock \url{https://doi.org/10.1063/1.5093988}

\bibitem[{{Grosjean} {et~al.}(2020){Grosjean}, {Wald}, {Sobarzo}, \&
  {Waitukaitis}}]{Grosjean2020}
{Grosjean}, G., {Wald}, S., {Sobarzo}, J.~C., \& {Waitukaitis}, S. 2020,
  Physical Review Materials, 4, 082602

\bibitem[{{Gundlach} \& {Blum}(2015)}]{Gundlach2015}
{Gundlach}, B., \& {Blum}, J. 2015, \apj, 798, 34

\bibitem[{{G{\"u}ttler} {et~al.}(2010){G{\"u}ttler}, {Blum}, {Zsom}, {Ormel},
  \& {Dullemond}}]{Guettler2010}
{G{\"u}ttler}, C., {Blum}, J., {Zsom}, A., {Ormel}, C.~W., \& {Dullemond},
  C.~P. 2010, \aap, 513, A56

\bibitem[{{Haeberle} {et~al.}(2018){Haeberle}, {Schella}, {Sperl},
  {Schr{\"o}ter}, \& {Born}}]{Haeberle2018}
{Haeberle}, J., {Schella}, A., {Sperl}, M., {Schr{\"o}ter}, M., \& {Born}, P.
  2018, Soft Matter, 14, 4987

\bibitem[{{Johansen} {et~al.}(2014){Johansen}, {Blum}, {Tanaka}, {Ormel},
  {Bizzarro}, \& {Rickman}}]{Johansen2014}
{Johansen}, A., {Blum}, J., {Tanaka}, H., {et~al.} 2014, in Protostars and
  Planets VI, ed. H.~{Beuther}, R.~S. {Klessen}, C.~P. {Dullemond}, \&
  T.~{Henning}, 547

\bibitem[{{Johansen} {et~al.}(2007){Johansen}, {Oishi}, {Mac Low}, {Klahr},
  {Henning}, \& {Youdin}}]{Johansen2007}
{Johansen}, A., {Oishi}, J.~S., {Mac Low}, M.-M., {et~al.} 2007, \nat, 448,
  1022

\bibitem[{{Johansen} \& {Youdin}(2007)}]{JohansenYoudin2007}
{Johansen}, A., \& {Youdin}, A. 2007, \apj, 662, 627

\bibitem[{{Jungmann} {et~al.}(2018){Jungmann}, {Steinpilz}, {Teiser}, \&
  {Wurm}}]{Jungmann2018}
{Jungmann}, F., {Steinpilz}, T., {Teiser}, J., \& {Wurm}, G. 2018, Journal of
  Physics Communications, 2, 095009

\bibitem[{{Jungmann} {et~al.}(2021){Jungmann}, {Bila}, {Kleinert},
  {M{\"o}lleken}, {M{\"o}ller}, {Schmidt}, {Schneider}, {Teiser}, {Utzat},
  {Volkenborn}, \& {Wurm}}]{Jungmann2021}
{Jungmann}, F., {Bila}, T., {Kleinert}, L., {et~al.} 2021, \icarus, 355, 114127

\bibitem[{Kaponig {et~al.}(2020)Kaponig, M{\"o}lleken, Tarasevitch, Utzat,
  Nienhaus, \& M{\"o}ller}]{Kaponig2020}
Kaponig, M., M{\"o}lleken, A., Tarasevitch, D., {et~al.} 2020, Journal of
  Electrostatics, 103, 103411

\bibitem[{{Klahr} \& {Schreiber}(2020)}]{Klahr2020}
{Klahr}, H., \& {Schreiber}, A. 2020, \apj, 901, 54

\bibitem[{{Kruss} {et~al.}(2016){Kruss}, {Demirci}, {Koester}, {Kelling}, \&
  {Wurm}}]{Kruss2016}
{Kruss}, M., {Demirci}, T., {Koester}, M., {Kelling}, T., \& {Wurm}, G. 2016,
  \apj, 827, 110

\bibitem[{{Kruss} {et~al.}(2017){Kruss}, {Teiser}, \& {Wurm}}]{Kruss2017}
{Kruss}, M., {Teiser}, J., \& {Wurm}, G. 2017, \aap, 600, A103

\bibitem[{{Kruss} \& {Wurm}(2018)}]{Kruss2018}
{Kruss}, M., \& {Wurm}, G. 2018, \apj, 869, 45

\bibitem[{{Kruss} \& {Wurm}(2020)}]{Kruss2020c}
---. 2020, The Planetary Science Journal, 1, 23

\bibitem[{{Lacks} \& {Mohan Sankaran}(2011)}]{Lacks2011}
{Lacks}, D.~J., \& {Mohan Sankaran}, R. 2011, Journal of Physics D Applied
  Physics, 44, 453001

\bibitem[{{Lee} {et~al.}(2018){Lee}, {James}, {Waitukaitis}, \&
  {Jaeger}}]{Lee2018}
{Lee}, V., {James}, N.~M., {Waitukaitis}, S.~R., \& {Jaeger}, H.~M. 2018,
  Physical Review Materials, 2, 035602

\bibitem[{{Lee} {et~al.}(2015){Lee}, {Waitukaitis}, {Miskin}, \&
  {Jaeger}}]{Lee2015}
{Lee}, V., {Waitukaitis}, S.~R., {Miskin}, M.~Z., \& {Jaeger}, H.~M. 2015,
  Nature Physics, 11, 733

\bibitem[{{Meisner} {et~al.}(2012){Meisner}, {Wurm}, \& {Teiser}}]{Meisner2012}
{Meisner}, T., {Wurm}, G., \& {Teiser}, J. 2012, \aap, 544, A138

\bibitem[{{Schaffer} {et~al.}(2018){Schaffer}, {Yang}, \&
  {Johansen}}]{Schaffer2018}
{Schaffer}, N., {Yang}, C.-C., \& {Johansen}, A. 2018, \aap, 618, A75

\bibitem[{Simon {et~al.}(2016)Simon, Armitage, Li, \& Youdin}]{Simon2016}
Simon, J.~B., Armitage, P.~J., Li, R., \& Youdin, A.~N. 2016, The Astrophysical
  Journal, 822, 55

\bibitem[{{Squire} \& {Hopkins}(2018)}]{Squire2018}
{Squire}, J., \& {Hopkins}, P.~F. 2018, MNRAS, 823

\bibitem[{{Steinpilz} {et~al.}(2020{\natexlab{a}}){Steinpilz}, {Joeris},
  {Jungmann}, {Wolf}, {Brendel}, {Teiser}, {Shinbrot}, \&
  {Wurm}}]{Steinpilz2020a}
{Steinpilz}, T., {Joeris}, K., {Jungmann}, F., {et~al.} 2020{\natexlab{a}},
  Nature Physics, 16, 225

\bibitem[{{Steinpilz} {et~al.}(2020{\natexlab{b}}){Steinpilz}, {Jungmann},
  {Joeris}, {Teiser}, \& {Wurm}}]{Steinpilz2020b}
{Steinpilz}, T., {Jungmann}, F., {Joeris}, K., {Teiser}, J., \& {Wurm}, G.
  2020{\natexlab{b}}, New Journal of Physics, 22, 093025

\bibitem[{{Weidenschilling} \& {Cuzzi}(1993)}]{Weidenschilling1993}
{Weidenschilling}, S.~J., \& {Cuzzi}, J.~N. 1993, in Protostars and Planets
  III, ed. E.~H. {Levy} \& J.~I. {Lunine}, 1031

\bibitem[{{Weidling} {et~al.}(2009){Weidling}, {G{\"u}ttler}, {Blum}, \&
  {Brauer}}]{Weidling2009}
{Weidling}, R., {G{\"u}ttler}, C., {Blum}, J., \& {Brauer}, F. 2009, \apj, 696,
  2036

\bibitem[{{Wurm} \& {Blum}(1998)}]{Wurm1998}
{Wurm}, G., \& {Blum}, J. 1998, \icarus, 132, 125

\bibitem[{{Wurm} {et~al.}(2019){Wurm}, {Schmidt}, {Steinpilz}, {Boden}, \&
  {Teiser}}]{Wurm2019}
{Wurm}, G., {Schmidt}, L., {Steinpilz}, T., {Boden}, L., \& {Teiser}, J. 2019,
  \icarus, 331, 103

\bibitem[{{Yang} {et~al.}(2017){Yang}, {Johansen}, \& {Carrera}}]{Yang2017}
{Yang}, C.~C., {Johansen}, A., \& {Carrera}, D. 2017, \aap, 606, A80

\bibitem[{Youdin \& Goodman(2005)}]{Youdin2005}
Youdin, A.~N., \& Goodman, J. 2005, The Astrophysical Journal, 620, 459

\bibitem[{{Zsom} {et~al.}(2010){Zsom}, {Ormel}, {G{\"u}ttler}, {Blum}, \&
  {Dullemond}}]{Zsom2010}
{Zsom}, A., {Ormel}, C.~W., {G{\"u}ttler}, C., {Blum}, J., \& {Dullemond},
  C.~P. 2010, \aap, 513, A57

\end{thebibliography}

\end{document}